\documentclass[aps,pra,twocolumn,showpacs]{revtex4}%
\usepackage{amssymb}
\usepackage{graphicx}
\usepackage{dcolumn}
\usepackage{bm}
\usepackage{amsmath}
\usepackage{soul,color}
\usepackage{textcomp}
\usepackage{times}
\usepackage{csquotes}
\usepackage{booktabs}

\usepackage[colorlinks,linkcolor=cBlue,citecolor=cBlue]{hyperref}
\usepackage{amsfonts} 
\usepackage{array}
\usepackage{longtable}
\providecommand{\U}[1]{\protect\rule{.1in}{.1in}}
\definecolor{cGreen}{RGB}{0,0,0}
\definecolor{cBlue}{RGB}{45,51,180}
\definecolor{cmagenta}{RGB}{205,0,100}

\begin{document}
	
	\title{From Ergodicity to Many-Body Localization in a One-Dimensional Interacting Non-Hermitian Stark System}
	
	\author{Jinghu Liu}
	\affiliation{Institute of Theoretical Physics and State Key Laboratory of Quantum Optics and Quantum Optics Devices, Shanxi University, Taiyuan 030006, China}
	
	\author{Zhihao Xu}
	\email{xuzhihao@sxu.edu.cn}
	\affiliation{Institute of Theoretical Physics and State Key Laboratory of Quantum Optics and Quantum Optics Devices, Shanxi University, Taiyuan 030006, China}
	\affiliation{Collaborative Innovation Center of Extreme Optics, Shanxi University, Taiyuan 030006, China}

	\begin{abstract}
		Recent studies on disorder-induced many-body localization (MBL) in non-Hermitian quantum systems have attracted great interest. However, the non-Hermitian disorder-free MBL still needs to be clarified. We consider a one-dimensional interacting Stark model with nonreciprocal hoppings having time-reversal symmetry, the properties of which are boundary dependent. Under periodic boundary conditions (PBCs), such a model exhibits three types of phase transitions: the real-complex transition of eigenenergies, the topological phase transition, and the non-Hermitian Stark MBL transition. The real-complex and topological phase transitions occur at the same point in the thermodynamic limit but do not coincide with the non-Hermitian Stark MBL transition, which is quite different from the non-Hermitian disordered cases. By the level statistics, the system transitions from the Ginibre ensemble (GE) to the Gaussian orthogonal ensemble (GOE) to the Possion ensemble with the increase of the linear tilt potential's strength. The real-complex transition of the eigenvalues is accompanied by the GE-to-GOE transition in the ergodic regime. Moreover, the second transition of the level statistics corresponds to the occurrence of non-Hermitian Stark MBL. We demonstrate that the non-Hermitian Stark MBL is robust and shares many similarities with disorder-induced MBL, which several existing characteristic quantities of the spectral statistics and eigenstate properties can confirm. The dynamical evolutions of the entanglement entropy and the density imbalance can distinguish the real-complex and Stark MBL transitions. Finally, we find that our system under open boundary conditions lacks a real-complex transition, and the transition of non-Hermitian Stark MBL is the same as that under PBCs.
		
	\end{abstract}
	
	\pacs{}
	\maketitle
	\section{Introduction}
	
	Many-body localization (MBL) has revolutionized our understanding of quantum systems by revealing the existence of robust localized states in disordered interacting systems \cite{D. Basko2006, C.R. Laumann2014, R. Nandkishore2015, J.A. Kall2014, S. Bera2015, L. Rademaker2016, V. Khemani2017-2, N. Mace2019, Y. Bar Lev2015, E. Bairey2017, K. S. C. Decker2020}. It provides an example of a quantum interacting system that results in the preservation of a nonthermal state \cite{Joshua M Deutsch2018, A. De Luca2013, Y. Bar Lev2014, D. J. Luitz2016, D. A. Abanin2019}. Much theoretical effort was invested in its unusual features, such as the connections between MBL transitions and the random matrix theory \cite{T. Guhr1998,Y. Y. Atas2013}, the logarithmic increase in entanglement entropy with time \cite{J. H. Bardarson2012, M. Serbyn2013}, area law entanglement entropy for eigenstates \cite{B. Bauer2013, M. Serbyn2016}, the persistent density imbalance \cite{Q. Guo2021,Q. Guo2021L}, the emergent integrability \cite{V. Ros2015,C. Bertoni2023}, and the response to external probes \cite{S. Gopalakrishnan2015,T. C. Berkelbach2010} and periodic driving \cite{A. Lazarides2015,L. D’Alessio2013,V. Khemani2016}. The experimental community has also devoted significant attention to this field, particularly since it provides pathways to implement quantum memories and quantum dynamical control \cite{M. Schreiber2015, P. Bordia2016, T. Kohlert2019, J. Smith2016, P. Roushan2017, Q. Guo2021}. Experimental realizations of MBL have been achieved for different platforms, including ultracold atoms \cite{M. Schreiber2015, P. Bordia2016}, trapped ions \cite{J. Smith2016}, and superconducting circuits \cite{P. Roushan2017, Q. Guo2021}.
	
	The disorder is not the only mechanism to realize MBL, which can be used to localize single-particle states \cite{N. F. Mott1967, A. De Luca2014}. Some studies have suggested that MBL may exist in a translationally invariant system, such as in a system with gauge invariance or multiple particle components \cite{W. De Roeck2014,M. Brenes2018}. Recently, this issue was approached in another way: the study of the so-called Wannier-Stark localization of a noninteracting system in a uniformly tilted lattice \cite{G. H. Wannier1962,G. H. Wannier1960,H. Fukuyama1973} From this, interacting systems with Wannier-Stark potentials exhibit MBL-like characteristics, named Stark MBL, which has attracted considerable theoretical and experimental focus \cite{M. Schulz2019, Y.-Y. Wang2021,S. R. Taylor2020, L.Zhang2021,W. Morong2021,E. van Nieuwenburg2019, P. Ribeiro2020,R. Yao2020,G. H. Wannier1962,G. H. Wannier1960,H. Fukuyama1973,T. M. Gunawardana2022,X.-P. Jiang2023}.
	On the other hand, traditional quantum mechanics is based on the postulate of Hermiticity, which assumes that Hermitian operators represent physical observables. This postulate ensures that the eigenvalues of these operators are real and the corresponding eigenvectors are orthogonal \cite{Bender2007, Y. Ashida2020}. However, in recent years, there has been growing interest in exploring non-Hermitian quantum mechanics to understand and describe a wide range of physical phenomena that cannot be captured within the framework of traditional Hermitian quantum mechanics, such as the non-Hermitian skin effect (NHSE) \cite{K. Zhang2020,K. Zhang2022}, boundary-dependent spectra \cite{Z. Ou2023}, the failure of the bulk-edge correspondence \cite{S. Yao2018,D. S. Borgnia2020}, and the non-Bloch band theory \cite{K. Yokomizo2019, Y.-C. Wang2022,Zhihao Xu2020}. Introducing non-Hermiticity into disorder systems has brought a new perspective on the localization features. According to the random matrix theory, the spectral statistics of non-Hermitian disorder systems display distinct features from the Hermitian ones \cite{R. Hamazaki2020,Lucas Sa2020, Antonio. M2022, J. Ginibre1965}. The interplay between the on-site random disorder and the nonreciprocal hopping, first proposed by the pioneering works of Hatano and Nelson, reveals a finite delocalization-localization transition accompanied by a real-complex transition of a single-particle spectrum and a topological phase transition point \cite{N. Hatano1996, N. Hatano1997, N. Hatano1998}. Furthermore, one can find that the random on-site potential can suppress the complex spectrum of an interacting Hatano-Nelson model with time-reversal symmetry (TRS) under periodic boundary conditions (PBCs) \cite{R. Hamazaki2019}, which exhibits a coincidence of the spectral transition with the non-Hermitian MBL transition and the topological phase transition \cite{L.-Z. Tang2021}. Such a triple-phase transition has also been detected in TRS quasiperiodic systems under PBCs with and without interactions \cite{L.-J. Zhai2020,Z. Gong2018, Q. Lin2022,S. Weidemann2022}. However, due to anomalous behavior under open boundary conditions (OBCs) for a non-Hermitian case that describes the localization occurring at one of the boundaries of non-Hermitian open lattices for a vast number of bulk modes, more and more studies have focused on the so-called NHSE, which can be understood by applying the non-Bloch band theory. More recently, the fate of skin modes in interacting fermionic and bosonic systems and the robustness of the NHSE on the localization features of many-body disordered systems have been investigated \cite{Y.-C. Wang2022}.
	
	Here, we approach the question of non-Hermitian MBL without disorder from a different point of view by introducing interactions into a one-dimensional (1D) nonreciprocal single-particle model subjected to Wannier-Stark potentials with TRS. The so-called non-Hermitian Stark MBL is robust, exhibiting many similarities to and differences from non-Hermitian disorder-induced MBL with TRS. Under PBCs, with the uniform force increase, the system's spectral statistics change from the Ginibre ensemble (GE) to the Gaussian orthogonal ensemble (GOE) to Poisson statistics \cite{R. Hamazaki2020,Lucas Sa2020, Antonio. M2022}. The first transition corresponds to the real-complex and topological phase transitions in the thermodynamic limit. Moreover, the second transition corresponds to the non-Hermitian Stark MBL transition. These two transitions do not coincide, which is quite different from non-Hermitian disordered cases. In Refs. \cite{R. Hamazaki2019,L.-J. Zhai2020,L.-Z. Tang2021}, the authors studied a non-Hermitian disordered system with a random or quasiperiodic modulated on-site potential. Under PBCs, both non-Hermitian disordered systems display a coincidence of the real-complex transition, the topological phase transition, and the MBL transition. The corresponding statistical distribution behavior transitions from the GE to the Poisson ensemble with the increase of the disorder amplitude. However, under OBCs, the system exhibits a non-Hermitian Stark MBL with a spectral transition from GOE to Poisson statistics. We also study the similarities and differences of non-Hermitian many-body Stark systems with PBCs and OBCs using the dynamical entanglement entropy and the evolution of the density imbalance.

	The rest of this paper is organized as follows. In Sec. II, we introduce a 1D nonreciprocal lattice model in the presence of a linear potential. In Sec. III, we discuss the spectral transition, the topological phase transition, the level statistics, the static entanglement entropy, the entanglement entropy with time, and the quench dynamics of the density imbalance under PBCs. In Sec. IV, we study the non-Hermitian Stark MBL transition under OBCs. In Sec. V, we display the phase diagrams under PBCs and OBCs. Finally, a conclusion is given in Sec. VI.  
	
	\section{Model}
	
	We consider a 1D interacting Stark model with nonreciprocal hoppings with TRS, which can be described by
	\begin{align}
		\hat{H} &  =J \sum_{j} \left(e^{g}\hat{c}_{j}^{\dag}\hat{c}_{j+1}+e^{-g}\hat{c}_{j+1}^{\dag}\hat{c}_{j}\right)+\sum_{j} W_{j}\left( \hat{n}_{j}-\frac{1}{2}\right)  \nonumber\\
		&  +V \sum_{j}\left( \hat{n}_{j}-\frac{1}{2}\right)\left(  \hat{n}_{j+1}-\frac{1}{2}\right), \label{eq1}
	\end{align}
	where $\hat{c}_{j}^{\dag}$ is the fermionic creation operator at site $j$, $\hat{n}_{j}=\hat{c}_{j}^{\dag}\hat{c}_{j}$ is the associated particle-number operator, $J$ is the nearest-neighbor hopping strength, and $V$ is the nearest-neighbor interaction strength. One can tune the parameter $g$ to control the non-Hermiticity of the system. To realize a non-Hermitian setup, one can continuously monitor quantum many-body systems and justify the non-Hermitian dynamics for individual quantum trajectories with no quantum jumps \cite{T. E. Lee2014,J. Dalibard1992,M. Nakagawa2020,Y. Ashida2020}. We emphasize that studying the non-Hermitian treatment for a many-body Stark model is nontrivial, quite different from the Hermitian case and the master-equation approach \cite{E. Levi2016}. To study the non-Hermitian Stark MBL transition, we consider the on-site potential energy, which can be written as \cite{G. H. Wannier1962,G. H. Wannier1960,H. Fukuyama1973}
	\begin{equation}
		W_{j}=-\gamma j+\alpha \left(\frac{j}{L-1}\right)^2, \label{eq2}
	\end{equation}
	where $\gamma$ is the linear tilt strength, $\alpha$ represents the strength of the curvature, and $L$ is the system size. We introduce a small value of $\alpha$ to weakly break translation invariance, so the curvature's effect is to lift the degeneracies and stabilize the localization. The factor of $1/(L-1)^2$ in Eq. (\ref{eq2}) is required to prevent the curvature from dominating the linear part in the thermodynamic limit. For a noninteracting Hermitian case with $V=0$ and $g=0$, the single-particle wave functions display Wannier-Stark localization \cite{G. H. Wannier1962,G. H. Wannier1960,H. Fukuyama1973}. With interaction, the Hermitian case has Stark MBL, which was observed in Refs. \cite{M. Schulz2019,Y.-Y. Wang2021,S. R. Taylor2020, W. Morong2021}. The MBL of the Stark model with weak disorder was discussed in Refs. \cite{L.Zhang2021,E. van Nieuwenburg2019}. However, the non-Hermitian case has been discussed less.
	
	This work focuses on the non-Hermitian interacting Stark model with system size $L$ at half filling and total particle number $N=L/2$. We apply the exact diagonalization method to study this non-Hermitian system with Hilbert space dimension $D=\left(	\begin{array}[c]{c}	L\\	L/2 \end{array}	\right)$ under PBCs and OBCs, separately. For convenience, we set $J=1$ as the unit of energy and choose $V=1$, $g=0.1$, and $\alpha=0.5$ for our discussion. 
	
	\section{Interacting non-Hermitian Stark model under periodic boundary conditions}
	
	\subsection{Real-complex transition of eigenvalues}	
	
	\begin{figure}[tbp]
		\begin{center}
			\includegraphics[width=.5 \textwidth]{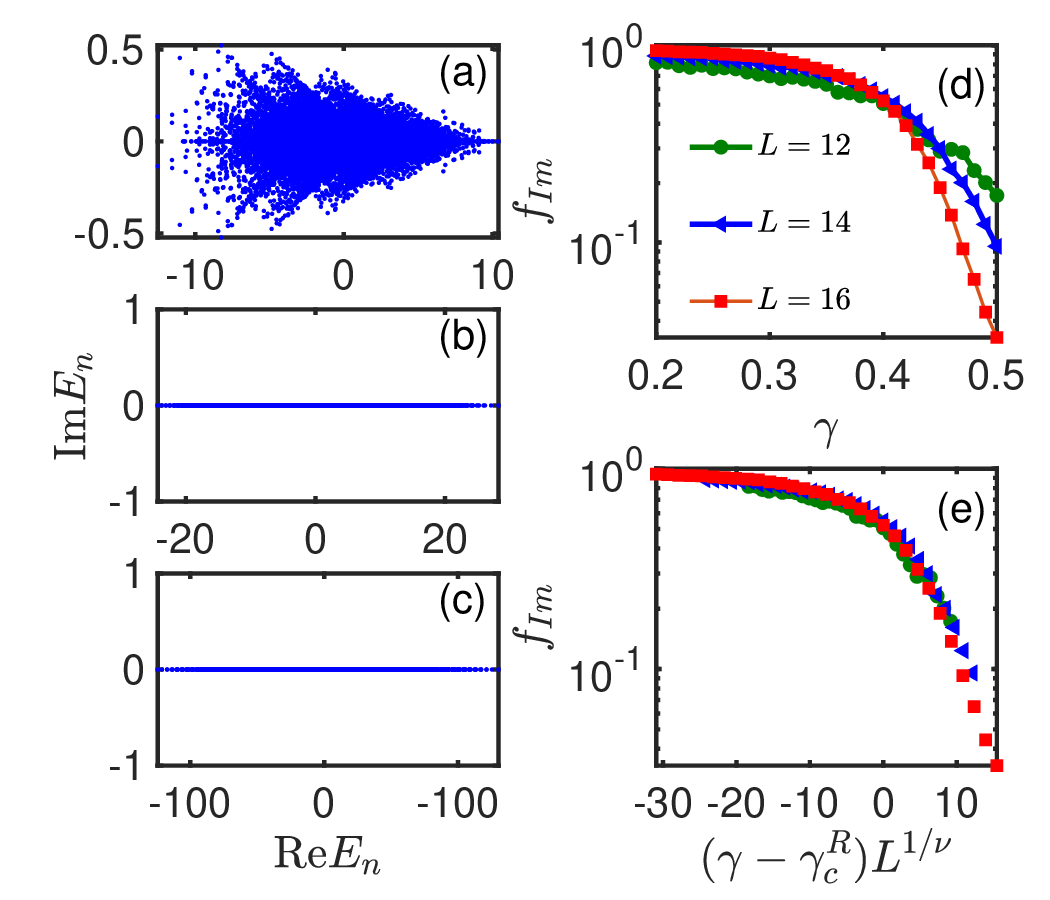}
		\end{center}
		\caption{(Color online) (a)\textendash(c) The eigenvalues of the Hamiltonian (\ref{eq1}) with $L=16$ for $\gamma =0.2$, $0.8$, and $4$, respectively. (d) The dependence of $f_{\mathrm{Im}}$ on $\gamma$ for different system sizes. As $L$ increases, $f_{\mathrm{Im}}$ increases for $\gamma \lesssim \gamma _{c}^{R}\left( \approx 0.4\right) $, and decreases for $\gamma \gtrsim \gamma _{c}^{R}$. (e) The critical scaling collapse of $f_{\mathrm{Im}}$ as a function of $\left(\gamma-\gamma _{c}^{R}\right) L^{1/\upsilon }$, with $\gamma_c^R=0.4$ and $\upsilon =0.55$. Here, we choose PBCs.}\label{Fig1}
	\end{figure}
	
	We first consider the spectral transition of the Hamiltonian (\ref{eq1}). Figures \ref{Fig1}(a)\textendash\ref{Fig1}(c) show the eigenvalues of the Hamiltonian (\ref{eq1}) with $L=16$ under PBCs for $\gamma=0.2$, $0.8$, and $4$, respectively. Due to the TRS, the energy spectrum is symmetric around the real axis. As the uniform force $\gamma$ increases, the eigenvalues with nonzero imaginary parts decrease. To quantitatively investigate the fraction of the complex eigenvalues, we define $f_{\mathrm{Im}}=D_{\mathrm{Im}}/D$ with a cut off $C=10^{-13}$ \cite{R. Hamazaki2019,L.-J. Zhai2020,K. Suthar2022}. When $|\mathrm{Im}{E}|\le C$, it is identified as a machine error. Here, $D_{\mathrm{Im}}$ is the number of the eigenvalues with nonzero imaginary parts. In Fig. \ref{Fig1}(d), we show $f_{\mathrm{Im}}$ as a function of $\gamma$ for different system sizes $L$. As $L$ increases, $f_{\mathrm{Im}}$ increases for $\gamma \le \gamma_c^{R} \approx 0.4$ and decreases for $\gamma \ge \gamma_c^{R}$. We further perform the finite-size scaling collapse of	$f_{\mathrm{Im}}$ by employing the ansatz $\left(\gamma-\gamma_{c}^{R}\right) L^{1/\upsilon}$ [Fig. \ref{Fig1}(e)], indicating a real-complex transition of many-body eigenvalues at $\gamma=\gamma_{c}^{R}$ with $\upsilon \approx 0.55$ in the thermodynamic limit. That means in the thermodynamic limit, almost all the eigenvalues are complex when $\gamma<\gamma_{c}^R$, and for $\gamma>\gamma_c^R$, the eigenvalues become real numbers. Similar results are found in a non-Hermitian many-body system with random or quasiperiodic on-site potentials.
	
	\subsection{Topological phase transition}
	
	\begin{figure}[tbp]
		\begin{center}
			\includegraphics[width=.50 \textwidth] {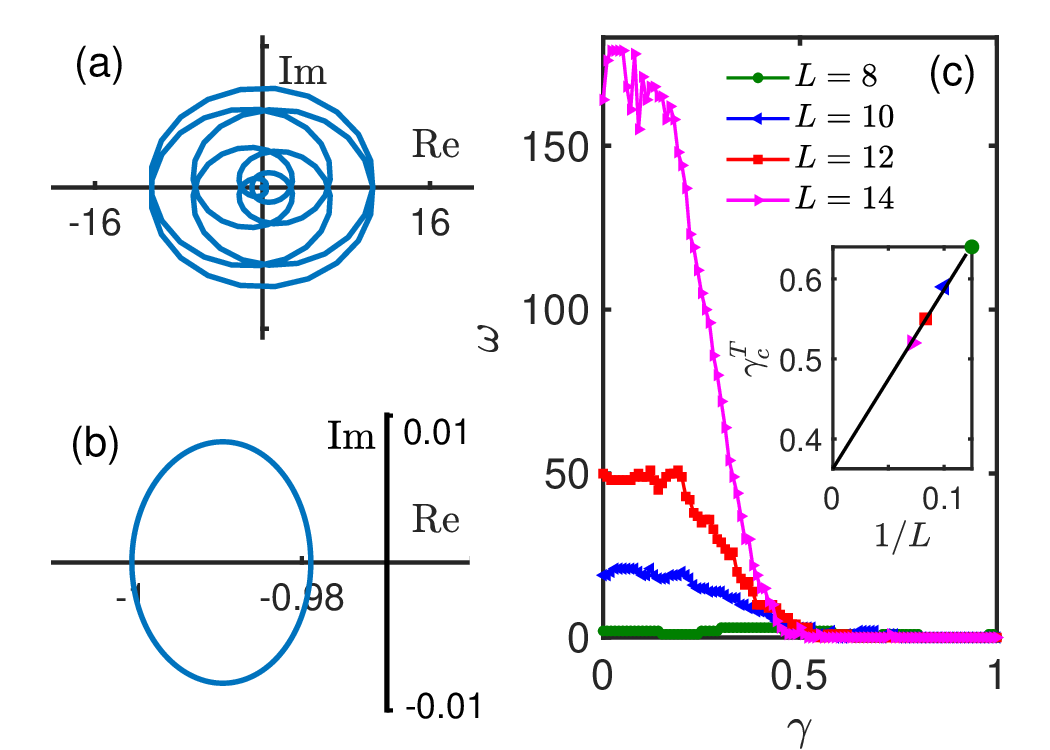}
		\end{center}
		\caption{(Color online) The dependence of $\det {[H(\Phi )]}/\left\vert \det{[H(0)]}\right\vert $ in the complex plane with $L=10$ for (a) $\gamma =0.4$ and (b)	$1.0$. (c) The winding number $\omega $ as a function of $\gamma$ with $L=8$, $10$, $12$, and $14$ for $V=1$. Inset: The topological phase transition $\gamma_c^{T}$ as a function of size $1/L$.}\label{Fig2}
	\end{figure}
	
	Unlike the Hermitian cases, our interacting non-Hermitian Stark system displays a topological phase transition in the complex energy plane \cite{Z. Gong2018,S. Longhi2019,L.-J. Zhai2020}. We introduce the winding number $\omega$ to characterize the topological feature of our system, which is given as follows \cite{Z. Gong2018,L.-J. Zhai2020}:
	\begin{equation}
		\omega=\int_{0}^{2\pi}\frac{d\Phi}{2\pi i}\partial_{\Phi}\ln\det\left[H\left(\Phi\right)-E_{B}\right], \label{eq3}
	\end{equation}
	where $H(\Phi)$ is the Hamiltonian (\ref{eq1}) under PBCs, the phase $\Phi$ is a magnetic flux penetrating through the center of the ring chain, and $E_B$ is the basis energy. The winding number $\omega$ counts the times the complex spectral trajectory encircles the chosen basis energy $E_B$ when the phase $\Phi$ rolls from $0$ to $2\pi$. One study demonstrated that the winding number does not depend on the choice of $E_B$ \cite{M. Arikawa2010}. In this case, we choose $E_B=0$. Since it is hard to directly apply Eq. (\ref{eq3}) for many-body systems, we alternatively use the $\Phi$ dependence of $\det{[H(\Phi)]}/\det{[H(0)]}$ to calculate the number of loops winding around $E_B$, which is equal to the winding number \cite{M. Arikawa2010}. Figures \ref{Fig2}(a) and \ref{Fig2}(b) show $\det{[H(\Phi)]}/\det{[H(0)]}$ in the complex plane with the rolling of the phase $\Phi$ from $0$ to $2\pi$ with $L=10$ for $\gamma=0.4$ and $1$, respectively. In Fig. \ref{Fig2}(a), we find that $\det{[H(\Phi)]}/\det{[H(0)]}$ draws a close loop around $E_B=0$ eight times, which corresponds to $\omega=8$. In contrast, $\det{[H(\Phi)]}/\det{[H(0)]}$ for $\gamma=1$ and $L=10$ in Fig. \ref{Fig2}(b) shows a close curve without surrounding $E_B=0$, corresponding to a topologically trivial case. Figure \ref{Fig2}(c) shows the winding number $\omega$ as a function of $\gamma$ for different $L$. As seen in Fig. \ref{Fig2}(c), the change in $\omega$ depends on the increase of $\gamma$, unlike in the single-particle Hermitian cases with only nearest-neighbor hoppings, in which the winding number is found to be $\omega=\pm 1$ for the topologically nontrivial phase. This implies that non-Hermitian many-body systems have much more complicated topological structures. On the other hand, the curves in Fig. \ref{Fig2}(c) show a transition from a topologically nontrivial phase with $\omega>0$ to a trivial one with $\omega=0$ at around $0.64$, $0.59$, $0.55$, and $0.52$ for $L=8$, $10$, $12$, and $14$, respectively. Note that the topological phase transition does not equal the real-complex transition in finite-size cases. However, complex eigenvalues are necessary to construct the close loops in the energy plane. The winding number defined in the complex plane by the gauge transformation is a collective indicator of eigenvalues being complex or real for the original Hamiltonian. The slight difference between $\gamma_c^R$ and $\gamma_c^T$ can be ascribed to the finite-size effect. A finite-size scaling of $\gamma_c^T$ with $1/L$ is shown in the inset of Fig. \ref{Fig2}(c). In the thermodynamic limit, the topological phase transition converges to the finite value $\gamma_c^T = 0.36\pm 0.04$ (the error represents the ${95\%}$ confidence interval), which is close to the real-complex transition point. Hence, we can conjecture that the transition points should coincide in the thermodynamic limit.

	\subsection{Non-Hermitian Stark many-body localization}
	
	\begin{figure}[tbp]
		\begin{center}
			\includegraphics[width=.5 \textwidth] {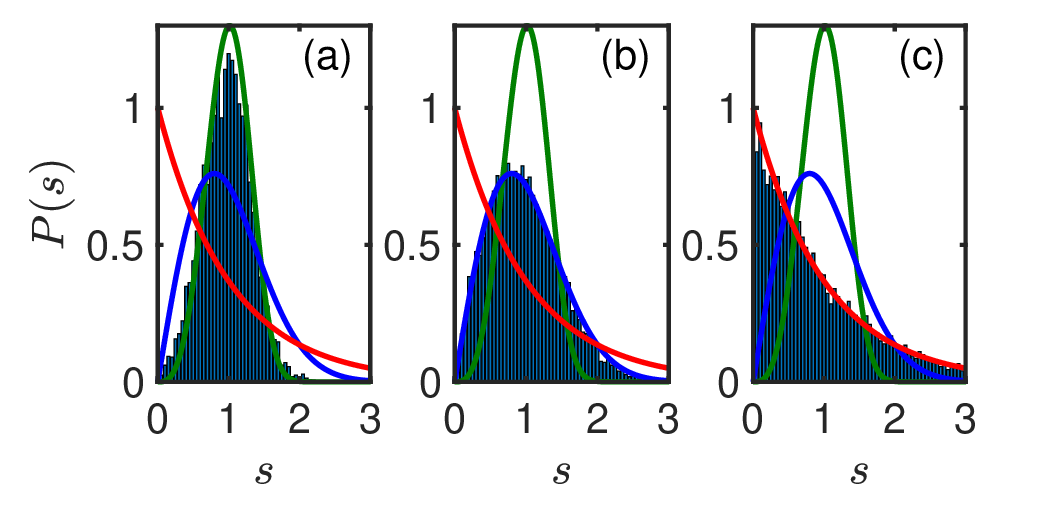}
		\end{center}
		\caption{(Color online) (a)\textendash(c) The unfolded nearest-level-spacing distribution of the Hamiltonian (\ref{eq1}) with $L=16$ for $\gamma=0.2$, $0.8$, and $4$, respectively. The green, blue, and red lines represent the Ginibre, GOE, and Poisson distributions, respectively. Here, we choose PBCs.}\label{Fig3}
	\end{figure}
	
	We next study an ergodicity-to-MBL transition in the interacting non-Hermitian Stark system. The level statistics is a powerful tool to diagnose the emergence of non-Hermitian MBL. In conventional non-Hermitian disorder systems with TRS, the ergodic phase follows the GE, and the MBL phase follows the real Poisson ensemble. We first consider the nearest-level-spacing distribution of unfolded eigenenergies in the complex plane \cite{T. Guhr1998,R. Hamazaki2019}. Here, the nearest level spacings for a given eigenvalue $E_n$ (before unfolding) are defined as $d_{1,n}=\min_{m}|E_n-E_m|$ in the complex energy plane, which is nonuniversal and system dependent. In order to compare theoretical predictions of random matrix theory with actually computed level sequences, one should perform an unfolding procedure. After a standard unfolding procedure, one can obtain the normalized level spacing $s$ with $\int_0^{\infty} p(s)ds =1$. The unfolding nearest level spacings $s_n=d_{1,n}\sqrt{\bar{\rho}_n}$, where $\bar{\rho}_{n} = \tilde{n}/(\pi d^2_{\tilde{n},n})$ is the local mean density; $\tilde{n}$ is sufficiently larger than unit, i.e., $\tilde{n} \approx 30$; and $d_{\tilde{n},n}$ is the $\tilde{n}$th nearest-neighbor distance from $E_n$. 
	
	For a small uniform force, the complex energy spectrum obeys the GE distribution $P_{\text{GE}}\left(  s\right)=cp\left(cs\right)$ shown in Fig. \ref{Fig3}(a) with $\gamma=0.2$, where \cite{R. Hamazaki2019}
	\begin{equation}
		p\left(s\right)=\lim_{N\rightarrow\infty}\left[  \prod_{n-1}^{N-1}e_{n}\left(s^{2}\right)e^{-s^{2}}\right]\sum_{n-1}^{N-1}\frac{2s^{2n+1}}{n!e_{n}\left(  s^{2}\right)}, \label{eq4}
	\end{equation}
	with $e_{n}\left(x\right)=\sum_{m=0}^{n}\frac{x^{m}}{m!}$ and $c=\int_{0}^{\infty}sp\left(s\right)ds=1.1429$, which matches the non-Hermitian random matrices in the AI symmetry class. When we further increase $\gamma$ beyond $\gamma_c^R$, the system's spectrum exhibits a real-complex transition, where the real eigenspectrum of a weak uniform force case follows the level statistics of the GOE. The level-spacing distribution of the GOE is \cite{R. Hamazaki2019}
	\begin{equation}
		P_{\text{GOE}}\left(s\right)  =\frac{\pi s}{2}e^{-\pi s^2/4}. \label{eq5}
	\end{equation}
	As shown in Fig. \ref{Fig3}(b) with $\gamma=0.8>\gamma_c^R$, the nearest-level-spacing distribution becomes a GOE case. For large enough $\gamma$, the system is immersed in the MBL phase with the real eigenspectrum, which is characterized by the real Poisson level distribution \cite{R. Hamazaki2019},
	\begin{equation}
		P_{\text{Pois}}\left(s\right)=e^{-s}. \label{eqp}
	\end{equation}	
	We take $\gamma=4$ as an example [Fig. \ref{Fig3}(c)]; the real eigenspectrum distribution becomes the Poissonian one. These results demonstrate that the non-Hermitian Stark system also has an MBL phase transition with the increase of $\gamma$. However, in the ergodic regime, the system undergoes a real-complex transition of eigenvalues, which leads to the nearest-neighbor level-spacing distribution from the GE distribution to the GOE distribution.
	
	\begin{figure}[tbp]
		\begin{center}
			\includegraphics[width=0.5 \textwidth] {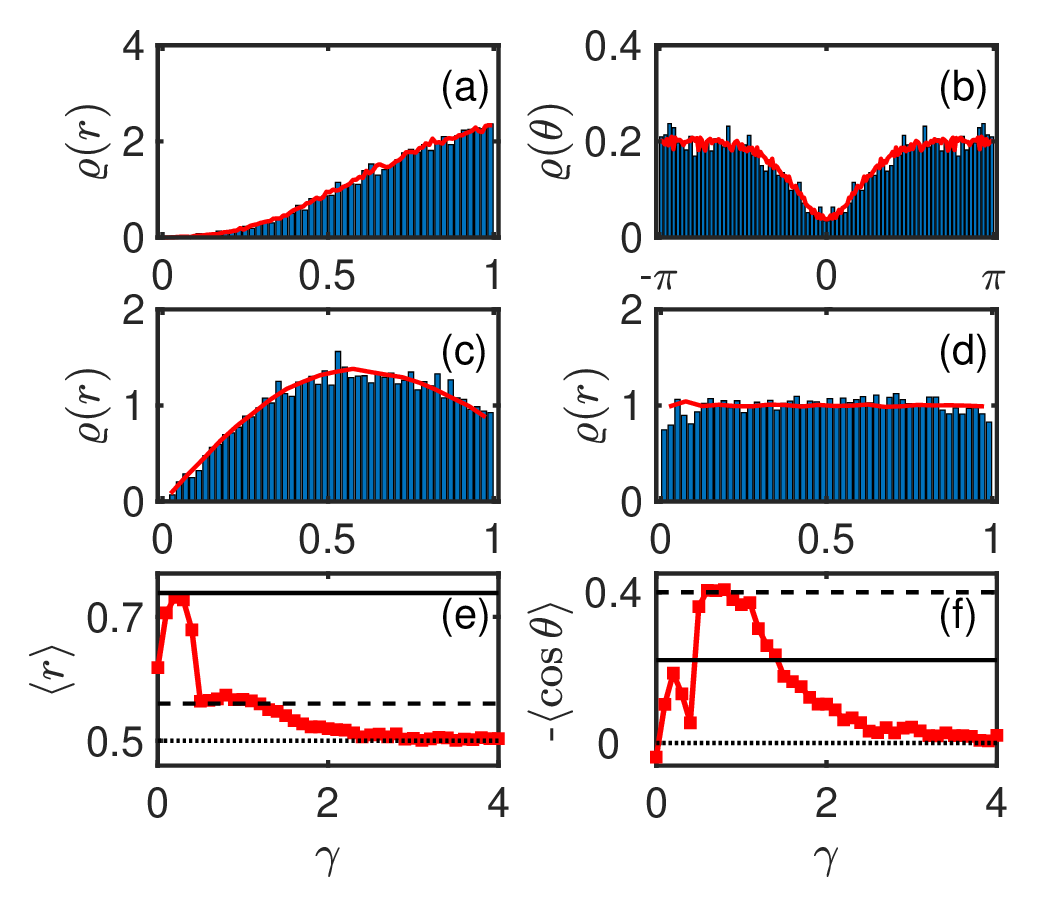}
		\end{center}
		\caption{(Color online) (a) and (b) The marginal distributions $\varrho(r)$ and $\varrho (\theta )$ with $\gamma =0.2$ for the complex energy spectrum. (c) and (d) The distributions $\varrho (r)$ with $\gamma =0.8$ and $4$ for the real energy spectrum, respectively. The solid red lines are obtained by calculating $\varrho(r)$ and $\varrho(\theta)$ of the $200\times 200$ symmetric random matrices with the corresponding random matrix ensembles averaged over $1000$ realizations. (e) The average magnitude $\left\langle r\right\rangle $ as a function of $\gamma$. The solid, dashed, and dotted lines represent $\langle r \rangle = 0.74$, $0.56$, and $0.5$, respectively. (f) $-\left\langle \cos \theta \right\rangle $ as a function of $\gamma$. The solid, dashed, and dotted lines denote $-\langle \cos{\theta} \rangle = 0.22$, $0.4$, and $0$, respectively. Here, we choose PBCs and $L=16$.}
		\label{Fig4}
	\end{figure}
	
	\begin{figure}[tbp]
		\begin{center}
			\includegraphics[width=0.5 \textwidth] {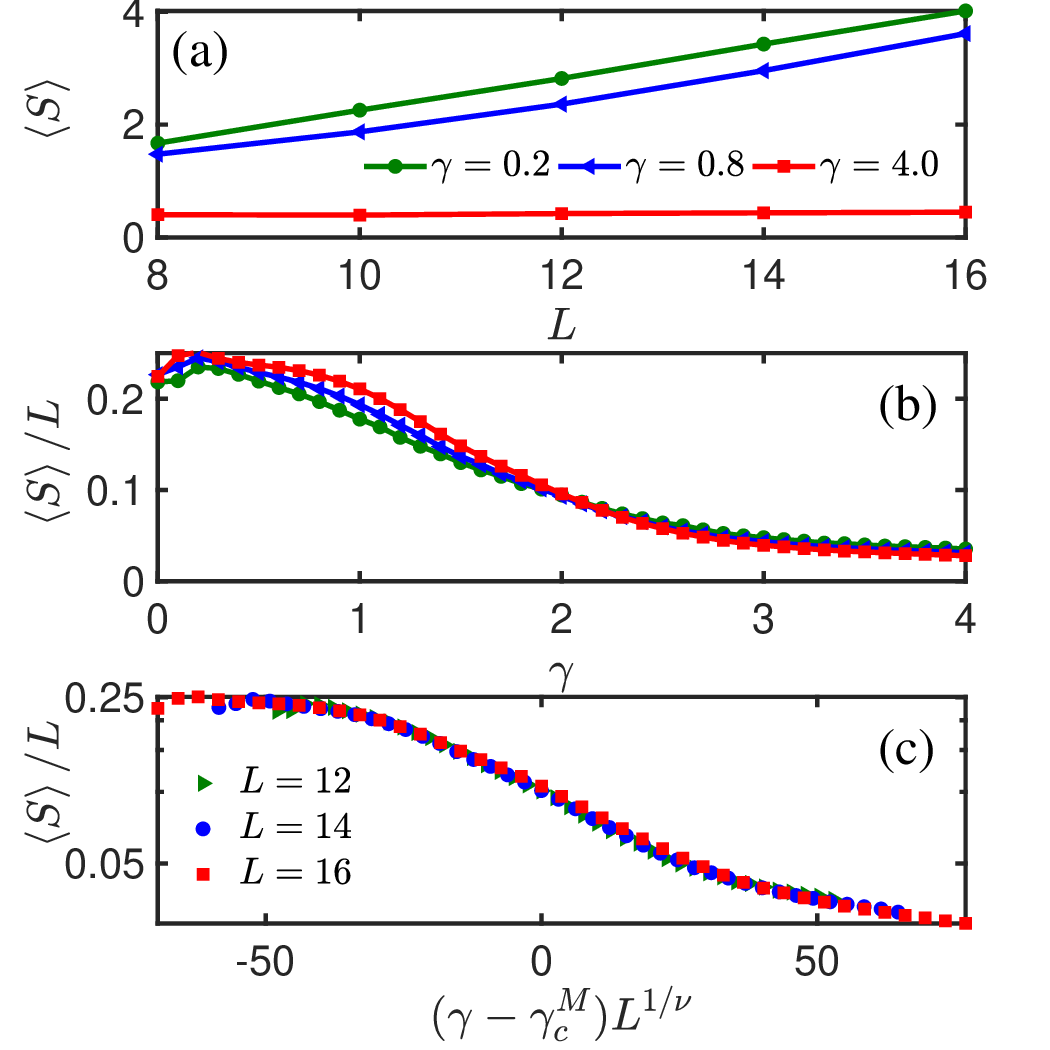}
		\end{center}
		\caption{(Color online) (a) The averaged half-chain entanglement entropy $\left\langle S\right\rangle $ as a function of $L$ for different $\gamma$. (b) $\left\langle S\right\rangle/L$ as a function of $\gamma$ for different $L$. (c) The critical scaling collapse of $\left\langle S\right\rangle/L$ as a function of $\left(\gamma-\gamma _{c}^{M}\right) L^{1/\mu }$, with $\gamma_c^M \approx 1.9$ and $\mu \approx 0.77$. Here, we choose PBCs.}
		\label{Fig5}
	\end{figure}

	We further consider that the complex spacing ratio (CSR) for the $n$th eigenvalue is defined as the ratio of complex differences, given by \cite{Lucas Sa2020}
	\begin{equation}
		\xi_{n}=\frac{E_{n}-E_{n}^{\mathrm{NN}}}{E_{n}-E_{n}^{\mathrm{NNN}}}=r_{n}e^{i\theta_{n}}, \label{eq6}
	\end{equation}
	where $E_{n}^{\mathrm{NN}}$ and $E_{n}^{\mathrm{NNN}}$ are the nearest and the next-nearest neighbors of the energy level $E_{n}$ in the complex plane, respectively. Note that $r_{n}$ and $\theta_{n}$ are the magnitude and the argument of the complex ratio $\xi_{n}$. The nearest-neighbor difference is nonuniversal and depends on the local density of states. In contrast, in the ratio $\xi_n$, the local density of state information is washed away. Hence, the CSR is a preferable diagnostic to detect an ergodicity-to-MBL transition. According to the definition of the CSR, $r_n \in [0,1]$, and $\theta_n\in (-\pi,\pi]$ $\forall n$. We focus on the radial and angular marginal distributions, denoted by  $\varrho(r)$ and  $\varrho(\theta)$, respectively. Both distributions $\varrho(r)$ and $\varrho(\theta)$ have distinct features for different random matrix ensembles. Figure \ref{Fig4} shows $\varrho(r)$ and $\varrho(\theta)$ for different $\gamma$ with $L=16$. For $\gamma=0.2$, the complex-valued spectrum of our system follows a GE distribution, in which the spectrum experiences level repulsion with the vanishing of $\varrho(r)$ for small $r$ [Fig. \ref{Fig4}(a)], and $\varrho(\theta)$ display a nonuniformity [Fig. \ref{Fig4}(b)]. When $\gamma>\gamma_c^R$, the eigenvalues become real, and the angular marginal distribution $\varrho(\theta)$ collapses to $\theta=0$ and $\pi$. When the system is localized in the ergodic regime following the GOE, $\varrho(r)$ displays behavior similar to that following the GE [see Fig. \ref{Fig4}(c), in which $\gamma=0.8$]. In Fig. \ref{Fig4}(d), in which $\gamma=4$, the system is immersed in the MBL phase with an uncorrelated energy level, and the corresponding $\varrho(r)$ is flat. The solid red lines in Fig. \ref{Fig4} were obtained by calculating $\varrho(r)$ and $\varrho(\theta)$ for the $200\times 200$ symmetric random matrices, with the corresponding random matrix ensembles averaged over $1000$ realizations, which are fit well with the interaction non-Hermitian Stark model. By detecting the average $\langle r\rangle$ and $-\langle \cos{\theta} \rangle$, we can obtain the phase transition information, where $\langle r \rangle =\int_0^1 dr r\varrho(r)$ and $-\langle \cos\theta\rangle =-\int_{-\pi}^{\pi} d\theta \cos\theta\varrho(\theta)$. We first consider $\langle r \rangle$ as a function of $\gamma$, as shown in Fig. \ref{Fig4}(e) with $L=16$. For a small $\gamma$ ($\gamma<\gamma_c^R$), $\langle r \rangle$ attains a constant value $\approx 0.74$ for the GE distribution \cite{K. Suthar2022,T. Peron2020,S. Ghosh2022,Lucas Sa2020}. When $\gamma$ is chosen as an intermediate value,  Fig. \ref{Fig4}(e) shows a transition to $\langle r \rangle \approx 0.56$, corresponding to the GOE \cite{T. Peron2020,K. Suthar2022}. This transition is consistent with the corresponding spectral transition. When $\gamma$ increases to the strong tilt limit, the system undergoes the $\gamma$-induced Stark MBL transition accompanied by a change in $\langle r \rangle$ from GOE to real Poisson statistics. For a real Poisson statistic, the non-Hermitian system has $\langle r \rangle \approx 0.5$ \cite{T. Peron2020,K. Suthar2022}. Likewise, the single-number signature of $-\langle \cos{\theta} \rangle$ can also distinguish the different phase regimes with different level distributions. For the system with the GE distribution $-\langle \cos{\theta} \rangle \approx 0.22$ \cite{K. Suthar2022,Lucas Sa2020,S. Ghosh2022}, which is shown in Fig. \ref{Fig4}(f) with $-\langle \cos{\theta} \rangle$ as a function of $\gamma$. As $\gamma$ increases, we find a dip in $-\langle \cos{\theta} \rangle$, which can demarcate the complex and real energy phases. In the GOE regime, $-\langle \cos{\theta} \rangle$ approaches a finite value of about $0.4$. Although the spectrum is real in this case, the real CSR $\xi_{n}\in\left[-1,1 \right]$, with the argument of $\xi_{n}$ being $\theta=0$ or $\pi$. We find that there are more $\xi_{n}$ with negative values than with positive values. By statistics, $-\langle\cos\theta\rangle$ approaches $0.4$ in the GOE regime. When the system goes into the Stark MBL phase with the Poisson statistics, $-\langle \cos{\theta} \rangle=0$. The main features of $-\langle \cos{\theta} \rangle$ agree with the behavior of $\langle r \rangle$. Notice that when $\gamma \to 0$, the values of $\langle r\rangle$ and $-\langle \cos{\theta} \rangle$ display a distinct deviation from the standard values in the GE regime. Observing the on-site potential (\ref{eq2}), we find that in the $\gamma \to 0$ limit, the quadratic term predominates. In the small-$\gamma$ limit, the symmetry axis of the potential function $j_o=\gamma (L-1)^2/(2\alpha)\in [0,L-1]$, which leads to the increase of the energy degeneracy and the deviation of the corresponding $\langle r\rangle$ and $-\langle \cos{\theta} \rangle$. To avoid the deviation, one can choose a larger $\gamma\geq2\alpha/\left(L-1\right)$ for discussion (see the Appendix for details).

	We use the static half-chain entanglement entropy $S_n=-\mathrm{Tr}[\rho_{L/2}^n\ln{\rho_{L/2}^n}]$ to exactly obtain the non-Hermitian Stark MBL transition point. Here, $\rho_{L/2}^n = \mathrm{Tr}_{L/2} [|E_n^{r}\rangle\langle E_n^r|]$, where $|E_n^r \rangle$ are normalized right eigenstates, i.e., $\langle E_n^r | E_n^{r} \rangle = 1$, and $\rho_{L/2}^n$ is the half-chain reduced density matrix obtained by tracing out half of the system. In Fig. \ref{Fig5}(a), we display the average entanglement entropy $\langle S \rangle$ averaged over all the right eigenstates as a function of $L$ with different $\gamma$. From Fig. \ref{Fig5}(a) where $\gamma=0.2$ and $0.8$, we can see that $\langle S \rangle$ follows a volume law in the ergodic phase. However, it decreases to a constant independent of $L$ in the deep MBL phase with $\gamma=4$, which fulfills an area law. Figure \ref{Fig5}(b) shows the $L$ dependence of $\langle S \rangle /L$ as a function of $\gamma$. In Fig. \ref{Fig5}(b), we can see that the average entanglement entropy $\langle S \rangle/L$ exhibits a crossover from the volume to area law as the non-Hermitian Stark MBL phase sets in around $\gamma_c^M \approx 1.9$. We confirm the critical scaling collapse as a function of $(\gamma-\gamma_c^{M})L^{1/\mu}$, which is shown in Fig. \ref{Fig5}(c), where $\mu=0.7$. These results show that the ergodic and stark MBL phases can be distinguished by entanglement entropy, even in non-Hermitian systems, like the Hermitian cases.
	
	According to our numerical calculation, there are three kinds of phase transitions: the real-complex transition of eigenvalues, the topological phase transition, and the non-Hermitian Stark MBL transition. The corresponding transition points $\gamma_c^{R}$ and $\gamma_c^T$ coincide in the thermodynamic limit but do not coincide with $\gamma_c^M$, which differs from the non-Hermitian cases with disordered on-site potentials.
	
	\subsection{Dynamical features}
	
	\begin{figure}[tbp]
		\begin{center}
			\includegraphics[width=0.5 \textwidth] {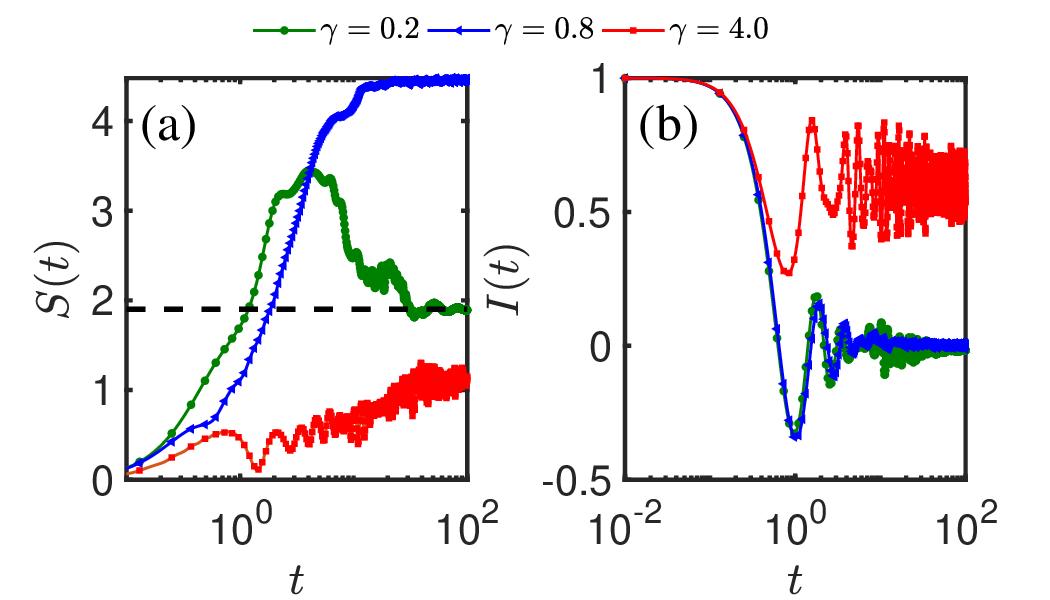}
		\end{center}
		\caption{(Color online) (a) Dynamics of the half-chain entanglement entropy $S\left(t\right)$ for different $\gamma$. (b) The dynamical evolution of the density imbalance $I(t)$ for different $\gamma$. Here, $L=16$ under PBCs, and the initial state is set as $\left\vert\psi _{0}\right\rangle=\left\vert 0101\cdots \right\rangle$.}\label{Fig6}
	\end{figure}
	
	In this section, we discuss the nonequilibrium time evolution of the non-Hermitian Stark system from the perspective of quantum trajectories without jump conditions for the continuously monitored system. We choose a given initial state $|\psi_0\rangle = |0101\cdots\rangle$ at $t=0$; the dynamical evolution can be encoded in the wave function $|\psi_t \rangle = e^{-i\hat{H}t}|\psi_0\rangle/\sqrt{\mathcal{N}}$ with the normalized coefficient $\mathcal{N}=\left\langle \psi_{0}\right\vert e^{iH^{\dag}t}e^{-iHt}\left\vert\psi_{0}\right\rangle$. With the help of the time-dependent wave function, many dynamical features can be detected.
	
	The dynamics of half-chain entanglement entropy can be defined as \cite{T. Orito2022}
	\begin{equation}
		S(t) = - \mathrm{Tr}[\rho_{L/2}(t)\ln \rho_{L/2}(t)], \label{eq7}
	\end{equation}
	where $\rho_{L/2}(t) = \mathrm{Tr}_{L/2}[|\psi_t\rangle \langle \psi_t |]$ is the reduced density matrix of $|\psi_t\rangle$. The time evolution of $S(t)$ for different $\gamma$ with $L=16$ is shown in Fig. \ref{Fig6}(a). For $\gamma=4$, the evolution of $S(t)$ displays logarithmic growth, which characterizes the Stark MBL. For the $\gamma=0.2$ and $0.8$ cases, the short-time evolution of $S(t)$ shows linear growth. However, $S(t)$ can decrease after $t \approx 10$ in the complex eigenvalue phase ($\gamma=0.2$) but remains a stable value in the real eigenenergy phase ($\gamma=0.8$). The results of the dynamical evolution of $S(t)$ imply that one can detect the occurrence of the Stark MBL using the short-time evolution of $S(t)$, and the long-time behavior of $S(t)$ signifies the real-complex transition. 
	
	We further investigate the dynamics of density imbalance for different $\gamma$, which is defined as \cite {Q. Guo2021, Q. Guo2021L,K. Suthar2022}
	\begin{equation}
		I(t) = \frac{N_e(t)-N_o(t)}{N_e(t)+N_o(t)}, \label{eq8}
	\end{equation}	
	where $N_e(t)=\sum_{\tilde{j}} \langle \psi_t| \hat{n}_{2\tilde{j}}|\psi_t \rangle$ and $N_o(t)=\sum_{\tilde{j}} \langle \psi_t| \hat{n}_{2\tilde{j}+1}|\psi_t \rangle$, with $\tilde{j}$ being an integer. Figure \ref{Fig6}(b) shows the dynamics of density imbalance $I(t)$ for different $\gamma$. In the ergodic phase, $I(t)$ relaxes to zero as time evolves, losing the memory of the initial information. For large $\gamma$, $I(t)$ remains nonzero at long times, which indicates that the system is localized in the Stark MBL phase.
	
	\section{non-Hermitian Stark many-body Localization under open boundary conditions}
	
	\begin{figure}[tbp]
		\begin{center}
			\includegraphics[width=0.5 \textwidth] {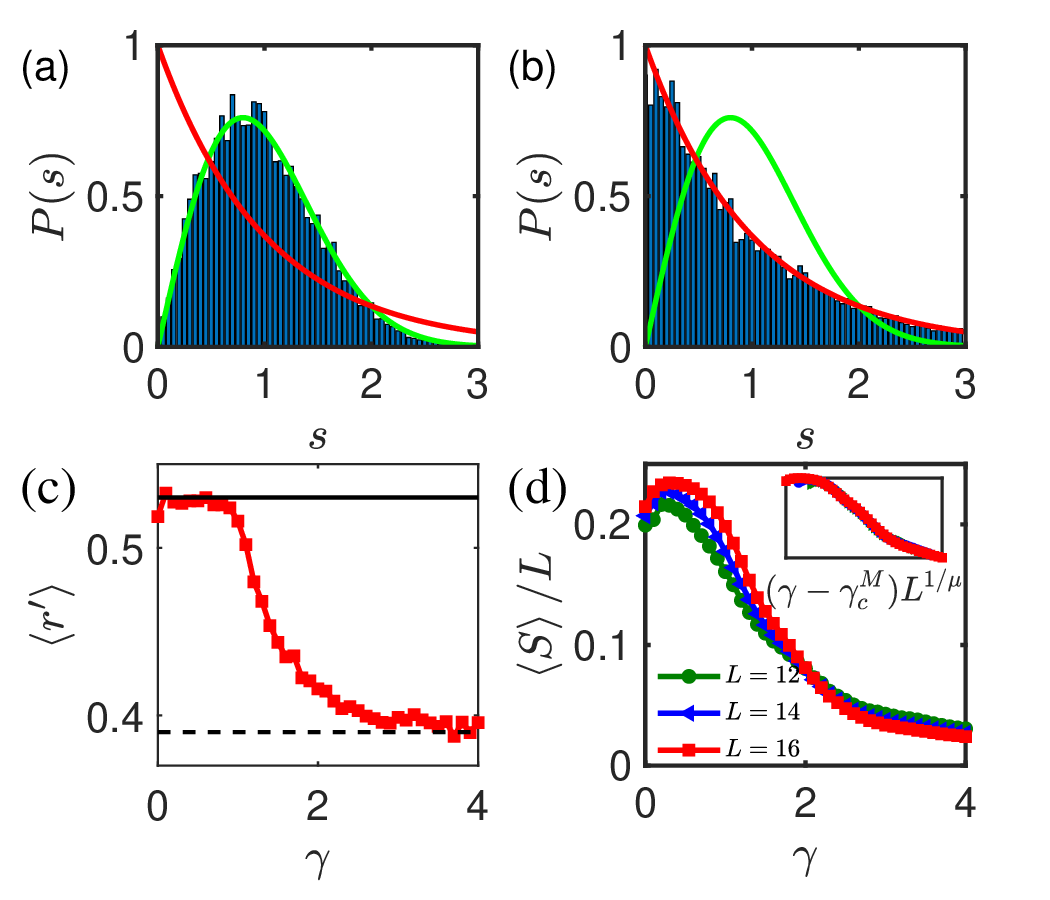}
		\end{center}
		\caption{(Color online) The unfolded nearest-level-spacing distribution of the Hamiltonian (\ref{eq1}) with $L=16$ for (a) $\gamma=0.2$ and (b) $\gamma =4$. The green and red lines represent the GOE and Poisson distributions. (c) The average ratio of adjacent energy gaps $\langle r^{\prime} \rangle$ as a function of $\gamma$ with $L=16$. The solid and	dashed lines correspond to the GOE and Poisson predictions, respectively. (d) $\left\langle S\right\rangle/L$ as a function of $\gamma$ for different $L$. Inset: The critical scaling collapse of $\left\langle S\right\rangle/L$ as a function of $\left(\gamma-\gamma _{c}^{M}\right) L^{-1/\mu }$, with $\gamma_c^M \approx 1.9$ and $\mu \approx 0.77$. Here, we choose OBCs.}  
		\label{Fig7}
	\end{figure}
	\begin{figure}[tbp]
		\begin{center}
			\includegraphics[width=0.5 \textwidth] {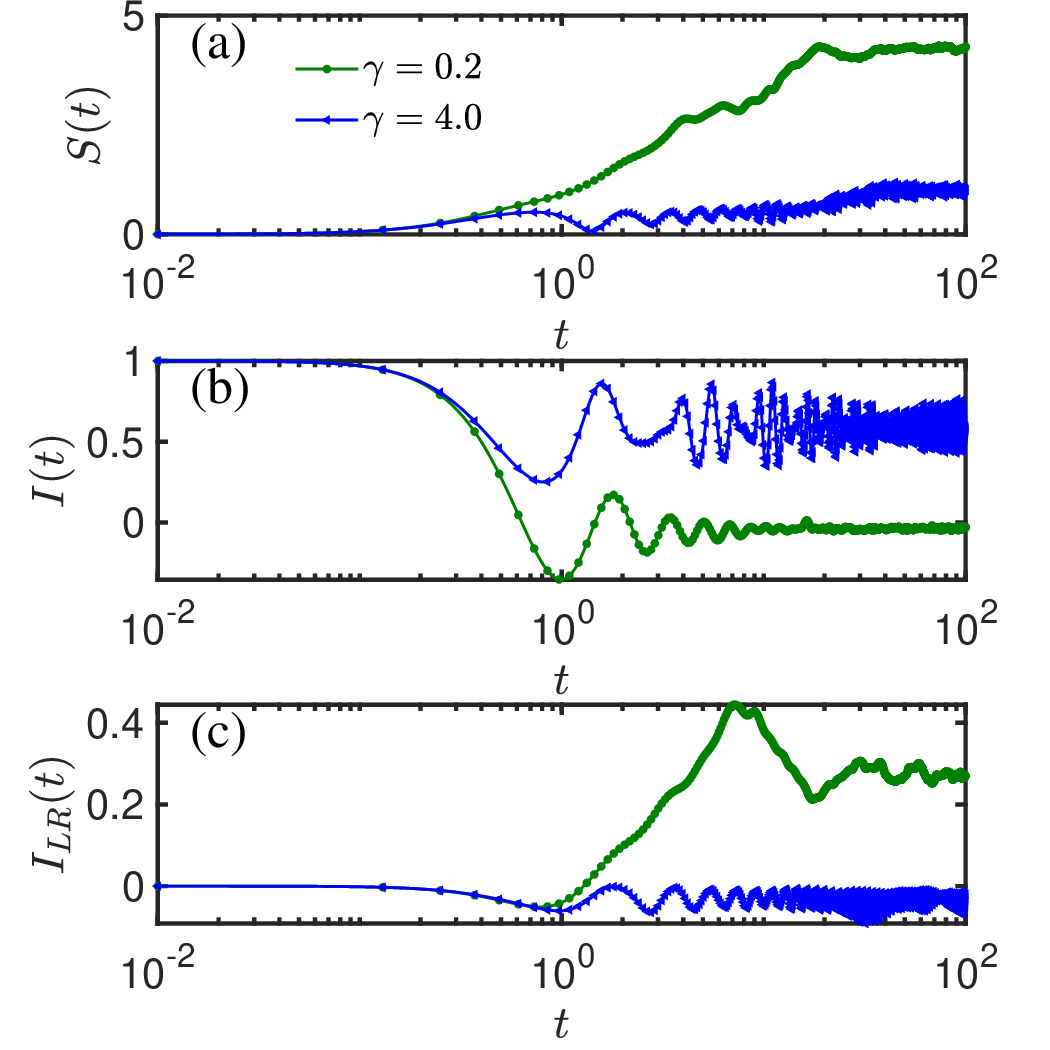}
		\end{center}
		\caption{(Color online) (a) Dynamics of the half-chain entanglement entropy $S\left(t\right)$ for different $\gamma$. (b) The dynamical evolution of the density imbalance $I(t)$ for different $\gamma$. (c) Dynamics of the left-right imbalance $I_{LR}\left( t\right) $ for different $\gamma$. Here, $L=16$ under OBCs, and the initial state is set as $\left\vert\psi _{0}\right\rangle=\left\vert 0101\cdots \right\rangle$. }
		\label{Fig8}
	\end{figure}
	
	The existence of the real spectrum under OBCs can be explained using an imaginary gauge transformation. Under OBCs, the nonreciprocal many-body Stark model with TRS can be mapped to a Hermitian model,
	\begin{align}
		\hat{H} &  =J \sum_{j} \left(\hat{c}_{j}^{\dag}\hat{c}_{j+1}+\hat{c}_{j+1}^{\dag}\hat{c}_{j}\right)+\sum_{j} W_{j}\left( \hat{n}_{j}-\frac{1}{2}\right)  \nonumber\\
		&  +V \sum_{j}\left( \hat{n}_{j}-\frac{1}{2}\right)\left(  \hat{n}_{j+1}-\frac{1}{2}\right), \label{eq1_real}
	\end{align} 
	by using a gauge transformation \cite{N. Hatano2021}, that is, $\hat{c}_{j}\rightarrow e^{-gj}\hat{c}_{j}$ and $\hat{c}_{j}^{\dag}\rightarrow e^{gj}\hat{c}_{j}^{\dag}$. Hence, under OBCs, the spectrum of the Hamiltonian (\ref{eq1}) is always real, and the real-complex transition of eigenvalues vanishes. In this section, we mainly discuss the Stark MBL transition under OBCs. 
	
	A well-established signature of the transition from the ergodic phase to the MBL phase is the level statistics of the spectrum. For our model under OBCs with the real spectrum in the delocalized and ergodic phase, we expect the GOE of the random matrix theory to be relevant. In contrast, the MBL phase with a real spectrum will lead to real Poisson level statistics. We perform a standard unfolding procedure of the real energy spectrum, obtaining level sequences of unit mean spacing. Figures \ref{Fig7}(a) and \ref{Fig7}(b) show the nearest-level-spacing distribution $P(s)$ of unfolded eigenvalues under OBCs with $L=16$ for $\gamma=0.2$ and $4$, respectively. For $\gamma=0.2$, $P(s)$ fits the GOE distribution described by Eq. (\ref{eq5}). However, when the system is localized in the MBL phase, the nearest-level-spacing distribution follows the real Poisson distribution in Eq. (\ref{eqp}), as shown in Fig. \ref{Fig7}(b) for $\gamma=4$. A simple indicator for a real spectrum to detect the MBL transition is the average ratio $\langle r^{\prime} \rangle$ between the smallest and largest adjacent energy gaps, given by \cite{Y. Y. Atas2013}
	\begin{equation}
		\langle r^{\prime} \rangle = \frac{\min\{ \delta_n^E,\delta_{n-1}^E\}}{\max\{\delta_n^E,\delta_{n-1}^E\}}, \label{eq9}
	\end{equation}
	where $\delta_n^{E} = E_n-E_{n-1}$ and $E_n$ is ordered in ascending order. In the ergodic (MBL) phase, $\langle r^{\prime} \rangle$ is close to the GOE value $\langle r^{\prime} \rangle \approx 0.53$ (the Poisson value $\langle r^{\prime}\rangle \approx 0.38$). As shown in Fig. \ref{Fig7}(c), we can see that $\langle r^{\prime} \rangle$ exhibits a clear transition from $0.53$ to $0.38$ with the increase of $\gamma$, which indicates the non-Hermitian Stark MBL also emerges under OBCs.
	
	We further determine the Stark MBL phase transition using the static half-chain entanglement entropy under OBCs. Figure \ref{Fig7}(d) shows the system-size dependence of $\langle S\rangle /L$ averaged over all the eigenstates for different $L$ as a function of $\gamma$ under OBCs. The static half-chain entanglement entropy exhibits a crossover from the volume to area law. We perform a finite-size critical collapse for $\langle S\rangle /L$, as shown in the inset of Fig. \ref{Fig7}(d). Our numerical results show that $\gamma_c^{M} \approx 1.9$ and $\mu \approx 0.77$ under OBCs are the same as those under PBCs.
	
	In Fig. \ref{Fig8}(a), we show the results for the dynamics of half-chain entanglement entropy $S(t)$ with the initial state $|0101\cdots \rangle$ under OBCs. For $\gamma>\gamma_c^M$ ($\gamma=4.0$), $S(t)$ grows as a logarithmic behavior and remains low in the long-time limit, indicating the emergence of Stark MBL. For $\gamma=0.2$, the short-time evolution of $S(t)$ shows linear growth. Unlike the PBC case, $S(t)$ remains a stable value without decreasing, which indicates that the spectrum of the system in the small-$\gamma$ case is real. 
	
	The NHSE is an iconic phenomenon which exhibits the localization of a large number of eigenstates at the boundaries under OBCs \cite{N. Okuma2023,Y.-C. Wang2023}. To distinguish the Stark MBL from the NHSE, we can utilize the dynamical evolution of the density imbalance $I(t)$ and the left-right imbalance \cite{F. Alsallom2022},
	\begin{equation}
		I_{LR}(t)=\sum_{j\le L/2}\langle \hat{n}_j\left( t \right)  \rangle -\sum_{j>L/2}\langle \hat{n}_j\left( t \right) \rangle, 
	\end{equation}
	with the even $L$, as shown in Figs. \ref{Fig8}(b) and \ref{Fig8}(c), respectively. Here, we choose the initial state $|0101\cdots \rangle$. When the system is localized in the ergodic phase $\gamma<\gamma_c^M$ exhibiting the NHSE, $I(t)$ decays quickly with $t$, and in the long-time limit, $I(t)\to 0$, and $I_{LR}(t)$ is finite. However, for $\gamma=4.0$, as shown in Fig. \ref{Fig8}(b), $I(t)$ displays a decay in the short-time limit; $I(t)$ remains finite, and $I_{LR}$ tends to zero in the long-time limit, which indicates that the system is localized in the Stark MBL phase without exhibiting the NHSE. 
	
	\section{Phase diagrams}
	\begin{figure}[tbp]
		\begin{center}
			\includegraphics[width=0.5 \textwidth]{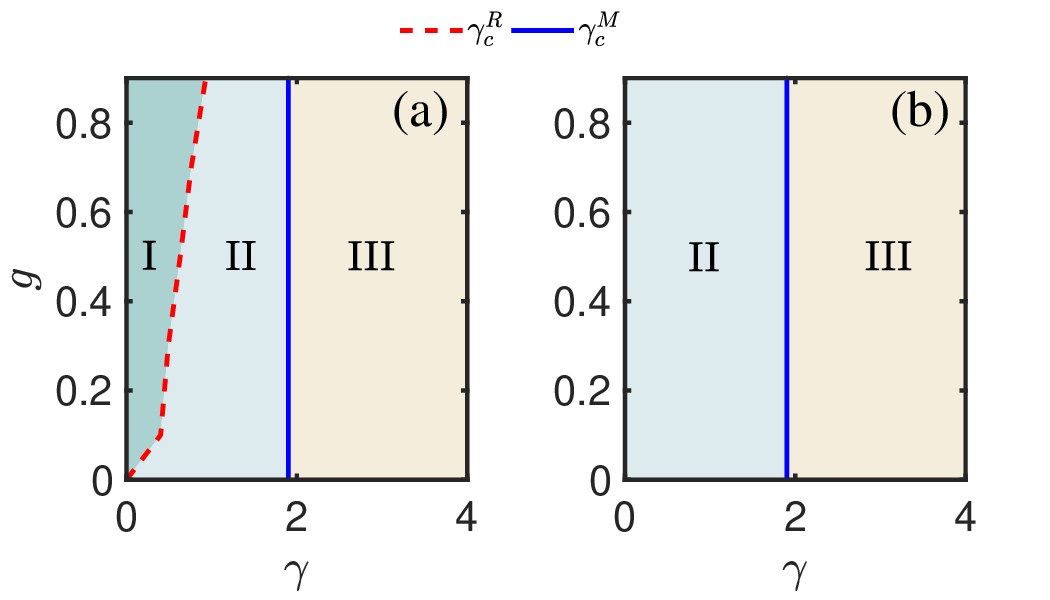}
		\end{center}
		\caption{(Color online) (a) Phase diagram for the non-Hermitian Stark model under PBCs, which contains three phases: the complex ergodic phase (phase I), the real ergodic phase (phase II), and the Stark MBL phase (phase III). The phase transition positions $\gamma_{c}^{R}$ and $\gamma_{c}^{M}$ are marked by the dashed and solid lines, respectively. (b) Phase diagram for the non-Hermitian Stark model under OBCs.} 
		\label{Fig9} 
	\end{figure}
	
	In this section, we observe the phase diagrams of Hamiltonian (\ref{eq1}) on the $g$-$\gamma$ plane under PBCs and OBCs, as illustrated in Figs. \ref{Fig9}(a) and \ref{Fig9}(b), respectively. As seen in Fig. \ref{Fig9}(a), there are three phases in this system under PBCs: phases I, II, and III correspond to the ergodic regime with the GE distribution, the ergodic regime with the GOE distribution, and the Stark MBL regime with the Poisson statistics, respectively. The phase transition positions $\gamma_{c}^{R}$ and $\gamma_{c}^{M}$ are marked by the dashed and solid lines, respectively. As $g$ increases, the regime with the complex spectrum gets larger. However, the transition of the Stark MBL is not sensitive to $g$. Under OBCs, as shown in Fig. \ref{Fig9}(b), there are two phases in the phase diagram due to the real spectrum for the whole parameter region. The Stark MBL transition points seem $g$ insensitive, i.e., $\gamma_{c}^{M}\approx1.9\pm0.1$, which is marked by the solid line in Fig. \ref{Fig9}(b).

	\section{Conclusions}
	
	In this paper, we first discussed the real-complex transition, topological	phase transition, and Stark MBL phase transition in an interacting non-Hermitian Stark model with TRS under PBCs. Unlike the non-Hermitian
	disordered cases with the random or quasiperiodic modulated on-site potential displaying the coincidence of the three transitions, our numerical results show that these three types of phase transition do not coincide. The level statistics show that the statistics change from the GE to the GOE to the Poisson ensemble under PBCs. The first transition corresponds to a spectral transition and the topological phase transition in the thermodynamic limit. Moreover, the second transition corresponds to an ergodicity-to-MBL transition. We also demonstrated that the non-Hermitian Stark MBL is robust and similar to disorder-induced MBL. The quench dynamics can corroborate the signature of the real-complex transition and the non-Hermitian Stark MBL. Finally, we studied the non-Hermitian Stark MBL under OBCs. Due to the real energy spectrum under OBCs, there is only one transition for the level statistics from the GOE to the Poisson ensemble, corresponding to the occurrence of non-Hermitian Stark MBL at $\gamma_{c}^{M} \approx 1.9$, the same as in the case under PBCs.
	
	Non-Hermitian systems with tunable nonreciprocal quantum transport were realized for ultracold atoms using dissipative Aharonov-Bohm rings in Refs. \cite{W. Gou2020,Q. Liang2022}, which has potential application in our non-Hermitian Stark model. By detecting the time-dependent atom population of each site, we could realize the measurement of the particle population over time in the non-Hermitian Stark model, allowing us to distinguish the NHSE and the Stark MBL using the dynamical quantities $I\left(t\right)$ and $I_{LR}\left(t\right)$. Besides relying on measuring dynamical quantities, analyzing the statistics of the energy spectrum is also a way to determine the Stark MBL. Fortunately, in a superconducting circuit, the many-body spectroscopy technique can retrieve the many-body eigenenergies and thereby provide information on the level statistics of the Hamiltonian \cite{D. A. Abanin2019,P. Roushan2017}. However, measuring energy spectra in non-Hermitian systems is challenging due to the complex energy levels. Further developments in energy spectrum measurement techniques are needed in these cases.
	
	\emph{Note added.} Recently, we came across a paper \cite{H.-Z.2023} in which the authors studied a similar problem and the main results of the emergence of the non-Hermitian Stark MBL were obtained. That paper emphasizes that the real-complex transition, the topological transition point, and the MBL transition are not identical in the interacting non-Hermitian Stark system under PBCs. However, we show the occurrence of the non-Hermitian Stark MBL without a real-complex transition in the energy spectrum under OBCs.
	
	\begin{acknowledgements}
		Z.X. is supported by the NSFC (Grant No. 12375016),the Fundamental Research Program of Shanxi Province, China (Grant No. 20210302123442), and the Open Project of Beijing National Laboratory for Condensed Matter Physics. This work is also supported by NSF for Shanxi Province (Grant No. 1331KSC).
		
	\end{acknowledgements}
	\appendix
	\section{Spectrum statistics for $\gamma \to 0$} \label{app_a}
	
	\begin{figure}[tbp]
		\begin{center}
			\includegraphics[width=0.5 \textwidth]{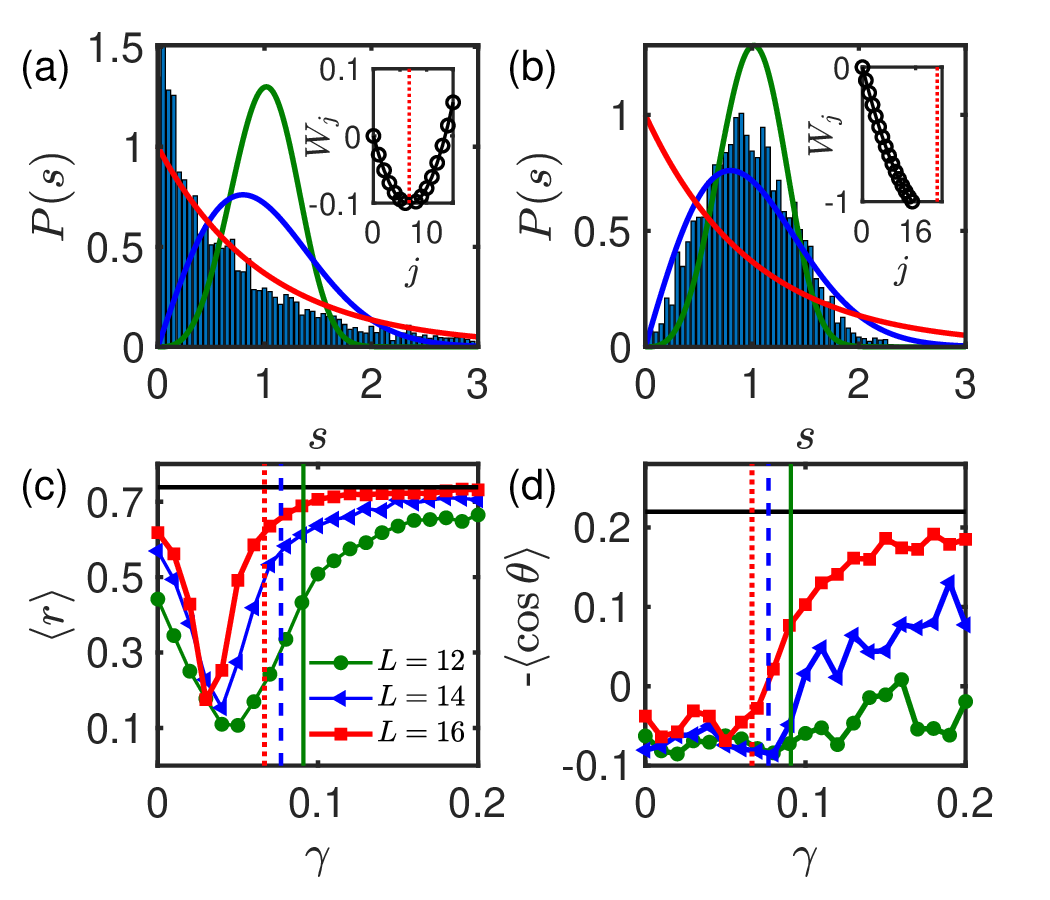}
		\end{center}
		\caption{(Color online) The unfolded nearest-level-spacing distribution of the Hamiltonian (\ref{eq1}) with $L=16$ for (a) $\gamma=0.03$ and (b) $\gamma =0.1$. The green, blue and red lines represent the Ginibre, GOE, and Poisson distributions, respectively. Inset: The on-site potential as a function of the lattice site. The vertical dotted line indicates the symmetry axis $j_o$. (c) $\left\langle r\right\rangle $ as a function of $\gamma$. The horizontal solid line represents $\langle r \rangle = 0.74$. (d) $-\left\langle \cos \theta \right\rangle $ as a function of $\gamma$. The horizontal solid line denotes $-\langle \cos{\theta} \rangle = 0.22$. The vertical solid, dashed, and dotted lines indicate $\gamma_s$ with $L=12$, $14$, and $16$, respectively. Here, we choose PBCs.} 	
		\label{Fig10} 
	\end{figure}
	
	In the main text, Figs. \ref{Fig4}(e) and \ref{Fig4}(f) show a distinct deviation from the standard values in the GE case for $\gamma\to 0$. The on-site potential (\ref{eq2}) displays a parabolic form. In the small-$\gamma$ limit, the quadratic part is predominant. We can see that a symmetry axis $j_{o}=\gamma(L-1)^2/(2\alpha)$ exists in the insets of Figs. \ref{Fig10}(a) and \ref{Fig10}(b), and it may induce the energy degeneracy. Figures \ref{Fig10}(a) and \ref{Fig10}(b) show the unfolded nearest-level-spacing distribution of the Hamiltonian (\ref{eq1}) with $L=16$ for $\gamma=0.03$ and $\gamma =0.1$, respectively. When $j_{o} \in [0,L-1]$, the corresponding level statistics show a Poisson-like distribution [see Fig. \ref{Fig10}(a)], where a peak emerges at $s \to 0$, indicating the existence of high degeneracy. When $j_{o}>L-1$, the level statistics restore the GE distribution [see Fig. \ref{Fig10}(b)], and the corresponding degeneracy breaks. To avoid the effect of the quadratic part, one can consider a larger $\gamma$, satisfying the condition: $\gamma\geq\gamma_s=2\alpha/\left(L-1\right)$, which can ensure the monotonicity of the on-site potential with the lattice site. To further verify the system size-dependent condition, we plot $\langle r \rangle$ and $-\langle \cos{\theta} \rangle$ as a function of $\gamma$ for different $L$ in Figs. \ref{Fig10}(c) and \ref{Fig10}(d), respectively. The different vertical lines correspond to $\gamma_s$ for different $L$. We can see that when $\gamma > \gamma_s$, $\langle r \rangle$ and $-\langle \cos{\theta} \rangle$ tend to the corresponding standard values of the GE distribution. Hence, for the small-$\gamma$ case, the deviation of the level statistics comes from the energy degeneracy. To avoid the deviation, one can consider the system-size-dependent condition $\gamma > \gamma_s$ for the calculation.

\end{document}